# Systematic evaluation of an atomic clock at $2 \times 10^{-18}$ total uncertainty

T.L. Nicholson[1,2], S.L. Campbell[1,2], R.B. Hutson[1,2], G.E. Marti[1,2], B.J. Bloom[1,2]*, R.L. McNally[1,2], W. Zhang[1,2], M.D. Barrett[1,3], M.S. Safronova[4,5], G.F. Strouse[6], W.L. Tew[6], and J. Ye[1,2]

## Abstract

The pursuit of better atomic clocks has advanced many research areas, providing better quantum state control, new insights in quantum science, tighter limits on fundamental constant variation, and improved tests of relativity. The record for the best stability and accuracy is currently held by optical lattice clocks. This work takes an important step towards realizing the full potential of a many-particle clock with a state-of-the-art stable laser. Our $^{87}$Sr optical lattice clock now achieves fractional stability of 2.2 × 10$^{-16}$ at 1 s. With this improved stability, we perform a new accuracy evaluation of our clock, reducing many systematic uncertainties that limited our previous measurements, such as those in the lattice ac Stark shift, the atoms' thermal environment, and the atomic response to room-temperature BBR. Our combined measurements have reduced the total uncertainty of the JILA Sr clock to 2.1 × 10$^{-18}$ in fractional frequency units.

## Introduction

Precise and accurate optical atomic clocks[1–5] have the potential to transform global timekeeping, enabling orders-of-magnitude improvements in measurement precision and sensor resolution for a wide range of scientific and technological applications. The pursuit of better atomic clocks has also had strong impact on many fundamental research areas, providing improved quantum state control[6,7], deeper insights in quantum science[8,9], tighter limits on fundamental constant variation[10,11], and enhanced sensitivity for tests of relativity[12]. Techniques developed for optical atomic clocks, such as advanced laser stabilization[13,14], coherent manipulation of atoms[15], and novel atom trapping schemes[16], have given rise to new research opportunities in quantum physics.

The continued advances in clock stability and accuracy go hand in hand. In an optical atomic clock, short-term stability originates from an ultrastable laser that serves as a local oscillator. Clock stability can be extended from seconds to hours by referencing the ultrastable laser to a high-quality-factor optical transition of an atom[17]. In this work, we use an ultrastable laser with 10 s coherence time, referenced at 60% duty cycle to thousands of strontium atoms in an optical lattice, to achieve a record fractional frequency stability of 2.2 × 10$^{-16}$ at 1 s.

Better clock stability allows for faster evaluations of systematic uncertainties and enables the discovery of new physical effects[18]. Here, we describe a set of innovations implemented to improve the accuracy of the $^{87}$Sr clock: an optical lattice with no measurable ac Stark shift at 1 ×

[1]JILA, National Institute of Standards and Technology and University of Colorado, Boulder, Colorado 80309-0440, USA. [2]Department of Physics, University of Colorado, Boulder, Colorado 80309-0390, USA. [3]Centre for Quantum Technologies, 3 Science Drive 2, Singapore 117543. [4]University of Delaware, Newark, Delaware 19716, USA. [5]Joint Quantum Institute, NIST and the University of Maryland, College Park, Maryland 20899, USA. [6]National Institute of Standards and Technology, Gaithersburg, MD 20899, USA. *Present address: Intel, Hillsboro, OR

$10^{-18}$, blackbody radiation (BBR) thermometry with millikelvin level accuracy, atomic structure measurements that characterize the atomic response to BBR, and active servo stabilization of electric and magnetic fields. With these developments, we achieve an overall systematic uncertainty of $2.1 \times 10^{-18}$, which is more than a threefold improvement over the previous best atomic clock[1]. This corresponds to a gravitational redshift for a height change of 2 cm on Earth.

## Results

### Clock stability

After preparing ultracold strontium atoms in an optical lattice (see Methods), we probe the $^1S_0 \rightarrow {}^3P_0$ 1 mHz clock transition with a 698 nm laser stabilized to 26 mHz[14]. The laser frequency offset from the clock transition is determined with Rabi spectroscopy, with lineshapes shown for 1 s and 4 s probe times in Fig. 1a. For longer probe times, atomic interactions affect the measured linewidth[18]. Here we use Fourier-limited probe times (≤1 s) to study the clock stability and systematics. The clock transition is probed once on each side of the resonance center; the difference in excited state fraction between these two measurements provides the error signal used to lock the laser to the clock transition.

Our clock stability at short averaging times is limited by the Dick effect[19]—aliased high frequency noise of the clock laser—that surpasses quantum projection noise[20] (QPN) with 2000 atoms. At long averaging times, the only mechanism that can limit the stability is drifting systematic shifts. We have demonstrated that after careful control of systematic effects, residual drifts did not affect clock stability at $2 \times 10^{-18}$ after thousands of seconds of averaging time[1]. Furthermore, stability data taken over the course of a month was robust and repeatable.

With long-term drift under control at the low $10^{-18}$ level, we can obtain a complete characterization of the clock stability with short-term stability measurements. Both the QPN and the Dick effect have been confirmed to be correctly determined with a self-comparison, which agrees with the measurement from a two-clock comparison[21]. A self-comparison approach compares two independent frequency locks operating on alternate experimental cycles[22]. Unlike synchronous stability[5,21], which is useful for systematic evaluations but which does not demonstrate how a system would perform as an independent frequency standard, a self-comparison reproduces the short-term stability of an independent clock.

Taking this approach, we use 1-s probe pulses to achieve the best independent clock stability of $2.2 \times 10^{-16}/\tau^{1/2}$, where $\tau$ is the averaging time in seconds (red solid line in Fig. 1b). This is consistent with our estimate of the Dick effect based on the known laser noise spectrum[14]. We now reach $1 \times 10^{-17}$ stability in less than 500 s, in contrast to the previous record of 1000 s (blue dashed line in Fig. 1b)[1,3,21].

This improved stability motivates the implementation of new strategies to reduce systematic uncertainties. Table I provides an uncertainty budget for our clock. We measure many of these uncertainties with lock-in detection, which involves modulating one parameter of our experiment between two values and recording the resulting frequency shift of the clock transition[1,22]. We present some of the important systematic shifts that are measured using lock-in detection, such as the lattice ac Stark and background dc Stark shifts. We also discuss two advances that reduce the BBR shift uncertainty: improved radiation thermometry and a direct measurement of the $^3D_1$ state lifetime to determine the atomic spectral response to BBR.

**Lattice ac Stark shift**

The lattice ac Stark shift is measured by performing lock-in detection of the frequency shift between different lattice intensities. Atoms are confined in an optical lattice with a tight trapping potential that eliminates Doppler and recoil shifts during clock spectroscopy. A magic wavelength optical trap[16,23,24] induces identical ac Stark shifts for the two clock states, making the clock transition frequency independent of the intensity of the optical trap.

The differential ac Stark shift of the two electronic clock states $\Delta\nu_{ac}$ is given by[25],

$$\Delta\nu_{ac} = U_0 \left\{ \Delta\kappa_s(f) + \Delta\kappa_v(f)m_F\xi\hat{k}\cdot\hat{B} + [3m_F^2 - F(F+1)]\left(3|\hat{e}\cdot\hat{B}|^2 - 1\right)\Delta\kappa_t(f)\right\}, \quad (1)$$

where $\hat{e}$ and $\hat{k}$ are the lattice polarization and propagation vectors, $\xi$ is the lattice polarization ellipticity (0 indicates linear polarization), $\hat{B}$ is the bias magnetic field direction which defines the quantization axis, $f$ is the lattice laser frequency, $U_0$ is the trap depth, and $\Delta\kappa_s$, $\Delta\kappa_v$, and $\Delta\kappa_t$ are the differential scalar, vector, and tensor shift coefficients, respectively. In our one-dimensional optical lattice geometry, we reduce sensitivity to drifts by aligning the bias magnetic field, the lattice light polarization, and the clock laser polarization ($\xi = 0$, $\hat{k}\cdot\hat{B} = 0$, $\hat{e}\cdot\hat{B} = 1$), as well as independently stabilizing the magnetic field[1]. To remove any residual vector Stark shift, we probe the stretched $m_F = \pm 9/2$ spin states and average their transition frequencies[26].

By varying both the lattice wavelength and $U_0$, we find the magic wavelength for $m_F = \pm 9/2$, where the scalar and tensor components of the differential Stark shift cancel[27,28] (Fig. 2b), and we operate our lattice there. The lattice laser is locked to an optical frequency comb that is referenced to the NIST Boulder hydrogen maser. Our operating wavelength is $c/(368.5544849(1)$ THz), where $c$ is the speed of light. Here we measure an ac Stark shift of (-1.3 ± 1.1) × 10$^{-18}$ for a trap depth of 12 μK, or 71 times the lattice photon recoil energy. At this lattice wavelength we do not observe a change in the clock frequency with lattice depth (Fig. 2a, open circles), in contrast to our previous measurement (Fig. 2a, open squares). We applied a linear fit to the data, because an *F*-test did not justify adding a term that is nonlinear in $U_0$ (see Methods).

Modulating the lattice depth changes the sample density, potentially adding a parasitic density shift. We account for this by employing a density shift cancellation[1] based on the experimentally verified relation that the density shift is proportional to $NU_0^{3/2}$. As $U_0$ is modulated, $N$ is correspondingly changed and monitored to ensure common-mode cancellation of the density shift.

## Dc Stark shift

The dc Stark shift is an important systematic effect that has been measured in lattice clocks[1,27,29]. Here, we demonstrate active control of the dc Stark along the axis that was found to have a measureable background field. Electrodes placed outside the vacuum chamber allow us to apply an external electric field and change its direction. Since the dc Stark shift is proportional to the square of the total electric field, a background field leads to a frequency difference when we reverse the applied field direction. This frequency difference, which is linearly proportional to the background electric field magnitude, serves as an error signal that is processed by a digital loop filter, which controls the electrode voltages to cancel the background Stark shift. This active servo, operated under a 0.1 Hz sampling rate, nulls the dc Stark shift with $1 \times 10^{-19}$ uncertainty.

## Radiation thermometry

The largest systematic uncertainty in our clock comes from the Stark shift $\Delta\nu_{BBR}$ due to the background BBR field[30,31]. $\Delta\nu_{BBR}$ can be approximated as,

$$\Delta\nu_{BBR} = \nu_{stat}\left(\frac{T}{T_0}\right)^4 + \nu_{dyn}\left(\frac{T}{T_0}\right)^6, \quad (2)$$

where $T$ is the ambient temperature, $T_0$ = 300 K, $\nu_{stat}$ and $\nu_{dyn}$ are the static and dynamic coefficients that describe the atomic response to ideal BBR, and higher order terms are negligible[32]. The static shift scales as $T^4$ because it is proportional to the total energy contained in the BBR electric field. The dynamic shift comes from coupling to atomic transitions out of the clock states that spectrally overlap with room temperature BBR, and is sensitive to deviations from an ideal BBR spectrum. Since $\nu_{stat}$ has already been accurately determined[32], the systematic uncertainty in $\Delta\nu_{BBR}$ comes from $\nu_{dyn}$ and $T$. While the dynamic term accounts for only 7% of the total BBR shift, uncertainty in $\nu_{dyn}$ is the dominant source of BBR shift uncertainty[1].

We measure the BBR environment of the atoms with thin-film platinum resistance thermometers[33] (PRTs), which are selected for good stability when thermally cycled over a test interval of 200 °C. Two PRTs (primary sensors) are painted black to increase radiative coupling and mounted to the ends of glass tubes sealed to vacuum flanges. Electrical feedthroughs allow for four-wire measurements (Fig. 3a). The PRTs are calibrated on their mounts at the NIST Sensor

Science Division temperature calibration facilities in Gaithersburg. Calibration is accomplished using Standard Platinum Resistance Thermometers traceable to the NIST ITS-90 temperature scale and water comparison bath. When there are temperature gradients across the mounting structures, heat that conducts from the flanges to the sensors (known as "immersion error") biases the BBR temperature measurements. To calibrate the bias, we embed a pair of secondary NIST-calibrated PRTs in the vacuum flanges (flange sensors) to measure these gradients. As a function of an applied gradient, we compare the primary sensor resistance in vacuum ($R_{vacuum}$), when the parasitic conductance is substantial, to the primary resistance in Helium ($R_{He}$), when the parasitic conductance is negligible (see Fig. 3b and Methods). After calibration, the sensors were returned under vacuum to JILA and installed in the clock vacuum chamber, where we observe that residual gradients in the clock chamber are very small and immersion errors are negligible.

Only in an inhomogeneous thermal environment do emissivities play a role in determining the dynamic BBR shift. Therefore, in order to predict the dynamic BBR shift correctly from the sensor resistance, we must ensure that the atoms are in a sufficiently thermal BBR environment. This is accomplished by surrounding the clock vacuum chamber with a BBR shield that achieves ≤ 1 K spatial temperature inhomogeneity (Fig. 3a). One sensor is moveable, and it measures a 1.5 mK temperature difference between the atom location and a retracted position 2.5 cm away. Accounting for our vacuum chamber emissivities and geometry, this small temperature gradient confirms a correction of less than $1 \times 10^{-19}$ to the clock uncertainty due to a non-thermal spectrum[1] (see Methods).

The final temperature uncertainties of the movable and fixed sensors are 5 mK and 11 mK, respectively. The agreement between the moveable and fixed sensors (Fig. 3c), which have markedly different immersion error coefficients (Fig. 3b), further ensures that the gradients in the clock chamber are small and supports the conclusion that no calibration shifts occurred during transport and installation. Using the movable sensor, with uncertainty summarized in Table II, we reach an uncertainty of $3 \times 10^{-19}$ in the static BBR shift. This approach allows us to operate the clock at room temperature while achieving a similar uncertainty to in-vacuum radiation-shielded lattice clocks at cryogenic[5] or room temperatures[34].

## $^3D_1$ decay rate

We now discuss our largest systematic uncertainty, which arises from the BBR dynamic coefficient $v_{dyn}$. The dominant source of uncertainty in $v_{dyn}$ comes from that of the oscillator strength of the 2.6 μm transition from the 5s5p $^3P_0$ clock state to the 5s4d $^3D_1$ state[30,32]. This is the only transition from a clock state that overlaps significantly in frequency with the room temperature BBR spectrum. According to Ref. 30, an accurate measurement of the 5s4d $^3D_1$ state lifetime $\tau_{3D1}$ will improve the $v_{dyn}$ accuracy. As shown in Fig. 4a, we first use our clock laser to drive the $^1S_0 \to {}^3P_0$ transition, and then use a 2.6 μm distributed-feedback laser to drive the $^3P_0 \to {}^3D_1$ transition with a 200 ns pulse. The atoms decay from the $^3D_1$ state into the $^3P$ manifold[35]. Those that decay into the $^3P_1$ state then decay to the $^1S_0$ state, spontaneously emitting a 689 nm photon that is collected on a photomultiplier. A photon counter time bins the data and we fit it to a double exponential function[35] to extract $\tau_{3D1}$ and the $^3P_1$ lifetime $\tau_{3P1}$ (Fig. 4b).

We use the fit function $y(t) = y_0 + A\{\exp[-(t-t_0)/\tau_{3P1}] - \exp[-(t-t_0)/\tau_{3D1}]\}$, where $t_0$ is the time offset, $y_0$ is the background counts, and $A$ is the amplitude. This functional form is valid after the 200 ns excitation pulse is extinguished as long as $t_0$ is a free fit parameter. Both an analytical model and a numerical simulation confirm that this functional form gives an unbiased fit. Another potential concern is density-dependent effects[35] such as radiation trapping and superradiance. However, as shown in Fig. 4c, we vary the density and observe no statistically significant density dependence of $\tau_{3D1}$. From our result of $\tau_{3D1}$ = (2.18 ± 0.01) μs we determine $\nu_{dyn}$ = (-148.7 ± 0.7) mHz, improving the uncertainty in $\nu_{dyn}$ by a factor of two and agreeing with Refs. 30,32. As shown in Table III, this measurement is limited by statistical error. The dynamic BBR uncertainty is reduced to $1.4 \times 10^{-18}$. We also improve the uncertainty of the $^3P_1$ lifetime by an order of magnitude, finding $\tau_{3P1}$ = (21.28 ± 0.03) μs.

Finally, we have greatly reduced the uncertainties in the first- and second-order Zeeman shifts and the probe Stark shift to the low $10^{-19}$ level or better (see Methods).

## Discussion

The current generation of stable lasers with >10 s coherence time and many-particle clocks have ushered in a new era of clock accuracy near the $1 \times 10^{-18}$ level. Even now, this coherence time has opened the possibility to eliminate the Dick effect by alternatively interrogating two separate atomic samples at >50% duty cycle with a single laser[36,37]. Soon, the next generation of ultrastable lasers will come online[13,38], with coherence times rivaling that of the 160 s natural lifetime of the Sr clock transition. The enhanced stability will not only bring clock accuracy to a new level, but also sets the stage for quantum metrology where quantum correlations will be harnessed to advance the frontier of measurement precision beyond the standard quantum limit[39–42].

**Acknowledgments:** We thank X. Zhang and M. Bishof for useful discussions and B. Bjork, L. Sonderhouse, and H. Green for technical assistance. This research is supported by the National Institute of Standards and Technology, Defense Advanced Research Projects Agency QuASAR Program, and NSF Physics Frontier Center at JILA. M.D.B acknowledges support from the JILA Visiting Fellows program. G.E.M. acknowledges support from the NIST Director's Office Fellowship.

**Author Information:** Reprints and permissions information is available at www.nature.com/reprints. The authors have no competing financial interest. Any mention of commercial products does not constitute an endorsement by NIST. Correspondence and requests for materials should be addressed to J.Y. (Ye@jila.colorado.edu).

**Figure 1. Single clock stability measured with a self-comparison.** a) A typical line scan associated with a 1 s interrogation time (open black circles). To explore the limit of coherence in our clock, we scan the clock transition with a 4 s interrogation time and more atoms (solid green squares). Here the linewidth and contrast are affected by the Fourier width and atomic interactions[18]. b) A new stability record (black circles, fit with red solid line) achieved by running with 1 s clock pulses and a 60% clock laser duty cycle for each preparation and measurement sequence. In contrast, the previous best independent clock stability[1,3,21] (blue dashed line) is $3.1 \times 10^{-16}/\sqrt{\tau}$. The error bars represent the 1σ uncertainty in the total deviation estimator, calculated assuming a white noise process, which is valid after the atomic servo attack time of ≈30 s.

**Figure 2. The ac Stark shift from the optical lattice.** a) Lattice ac Stark shift measurements, as a function of the differential trap depth $\Delta U$ (in units of lattice photon recoil energy), for the current evaluation (red circles) and our previous evaluation (blue squares). The lattice frequency for the new evaluation is 172.4 MHz lower than that of the previous evaluation. We determine the magic wavelength in our experimental configuration so that our trapping potential is independent of the electronic state ($^1S_0$ or $^3P_0$) for $m_F = \pm 9/2$. Our current evaluation thus achieves the smallest reported lattice ac Stark shift of (-1.3 ± 1.1) × 10$^{-18}$. Error bars represent 1σ uncertainties (calculated as described in Methods). b) The calculated lattice ac Stark shift $\Delta \nu_{ac}$ at the magic wavelength, plotted for different spin states. The trapping potential is independent of the electronic states when the scalar shift and the tensor shift cancel for $m_F = \pm 9/2$.

**Figure 3. Radiation thermometry in the JILA Sr clock.** a) Mounted radiation thermometers inside Sr clock chamber, surrounded by a BBR shield enclosure. Two thin film platinum resistance thermometers (PRTs) are mounted on glass tubes that are affixed to mini vacuum flanges. One sensor is fixed while the other can be translated to measure at the center of the vacuum chamber or, during normal clock operation, 2.5 cm from the center. The BBR shield is used for thermalization, minimizing temperature gradients and enabling passive temperature stabilization. b) The sensor calibration at the NIST Sensor Science Division. First, the sensor resistance is calibrated to the ITS-90 temperature scale under a He exchange gas ($R_{He}$). To calibrate the sensor resistance in vacuum ($R_{vacuum}$), we measure $R_{vacuum}$ - $R_{He}$ as a function of the temperature difference between the flange ($T_{flange}$) and primary ($T_{primary}$) sensors. The slopes quantify the immersion error coefficients, which are markedly different between the two sensors. However, we find negligible immersion errors in the BBR-shielded clock chamber. c) A long-term record of the temperature and total BBR shift (upper plot), and the temperature difference (lower plot) measured by the two primary sensors. While temperature fluctuations are within a few hundred mK, the sensor temperature difference (black line) is well within the combined uncertainty of both sensors (shown as the gray 1σ confidence band), which indicates that no calibration shifts occurred during shipping and installation.

**Figure 4. The measurement of the $^3D_1$ decay rate.** a) The electronic states used for the decay rate measurement. First we drive the clock transition, and then we use a 200 ns laser pulse to drive the 2.6 μm $^3P_0 \to {}^3D_1$ transition. The $^3D_1$ state decays into the $^3P$ manifold with the branching ratios depicted in the panel. Photons from $^3P_1 \to {}^1S_0$ are collected by a photomultiplier tube (PMT). b) The sum of photon counts for 8 million decay events (black dots), fit with the function $y(t) = y_0 + A\{\exp[-(t-t_0)/\tau_{3P1}] - \exp[-(t-t_0)/\tau_{3D1}]\}$ (red curve). Data when the pulse is on is excluded to ensure an unbiased fit. The inset is the error ellipse for the fits of $\tau_{3D1}$ and $\tau_{3P1}$. c) Lifetime versus atom number. Comparing a constant model of this data to a model that is linear in density (the first-order correction for a density-dependent effect) using an *F*-test, we find no statistically-significant lifetime dependence on density. Error bars represent 1σ fit uncertainties (see Methods).

**Table I**

| Effect | Shift (×10$^{-18}$) | Uncertainty (×10$^{-18}$) |
|---|---|---|
| Lattice Stark | -1.3 | 1.1 |
| BBR static | -4562.1 | 0.3 |
| BBR dynamic | -305.3 | 1.4 |
| dc Stark | 0.0 | 0.1 |
| Probe Stark | 0.0 | 0.0 |
| 1$^{st}$-order Zeeman | -0.2 | 0.2 |
| 2$^{nd}$-order Zeeman | -51.7 | 0.3 |
| Density | -3.5 | 0.4 |
| Line pulling + tunneling | 0.0 | <0.1 |
| 2$^{nd}$-order Doppler | 0.0 | <0.1 |
| Background gas | 0.0 | <0.6 |
| Servo offset | -0.5 | 0.4 |
| AOM phase chirp | 0.6 | 0.4 |
| **Total** | **-4924.0** | **2.1** |

**Table I. Clock uncertainty budget.** Descriptions of these effects can be found in the main text and in Methods. The BBR static and dynamic shifts are calculated for the ambient temperature of 20.6 °C (Fig. 3c). The shifts and their corresponding $1\sigma$ uncertainties are quoted in fractional frequency units. The statistical uncertainties for each effect are inflated by the square root of the reduced chi-square statistic, $\chi^2_{red}$, when $\chi^2_{red} > 1$. Typical values of $\chi^2_{red}$ are between 1 and 1.5. Statistical uncertainties are summed in quadrature with the systematic uncertainties for each effect. The only significant uncertainties that are not based on measurement statistics are the BBR shift and the background gas correction.

**Table II**

| Effect | Uncertainty (mK) |
|---|---|
| Bath non-uniformity | 1.0 |
| Bath SPRT calibration | 1.0 |
| Bath temp. stability | 1.0 |
| Sensor self-heating | 0.5 |
| Electrical errors | 0.07 |
| Sensor translation | 0.03 |
| Thermal cycling | 2.0 |
| Calibration coefficients | 4.5 |
| **Total** | **5.2** |

**Table II. Radiation thermometer calibration uncertainty for the moveable sensor.** This sensor provides the temperature measurements used in the BBR shift correction for the clock. Each entry is a $1\sigma$ uncertainty. The bath calibration is monitored with two standard platinum resistance thermometers (SPRTs). See Methods for a description of each effect. The dominant uncertainty comes from a fit of ($R_{vacuum}$ - $R_{He}$) as a function of ($T_{flange}$ - $T_{primary}$) (Fig. 3b).

**Table III**

| Effect | Uncertainty (ns) |
|---|---|
| Fit uncertainty | 10 |
| Hyperfine correction | <0.1 |
| Finite pulse duration | <0.1 |
| Stray laser light | <0.01 |
| BBR contamination | <0.01 |
| Photon counter timing | 0.4 |
| **Total** | **10** |

**Table III. Uncertainty budget for $^3D_1$ decay rate.** Each entry is a $1\sigma$ uncertainty. Here the statistical fit uncertainty dominates the total uncertainty of this measurement. See Methods for a description of each effect. We do not measure a density-dependent effect within our statistics.

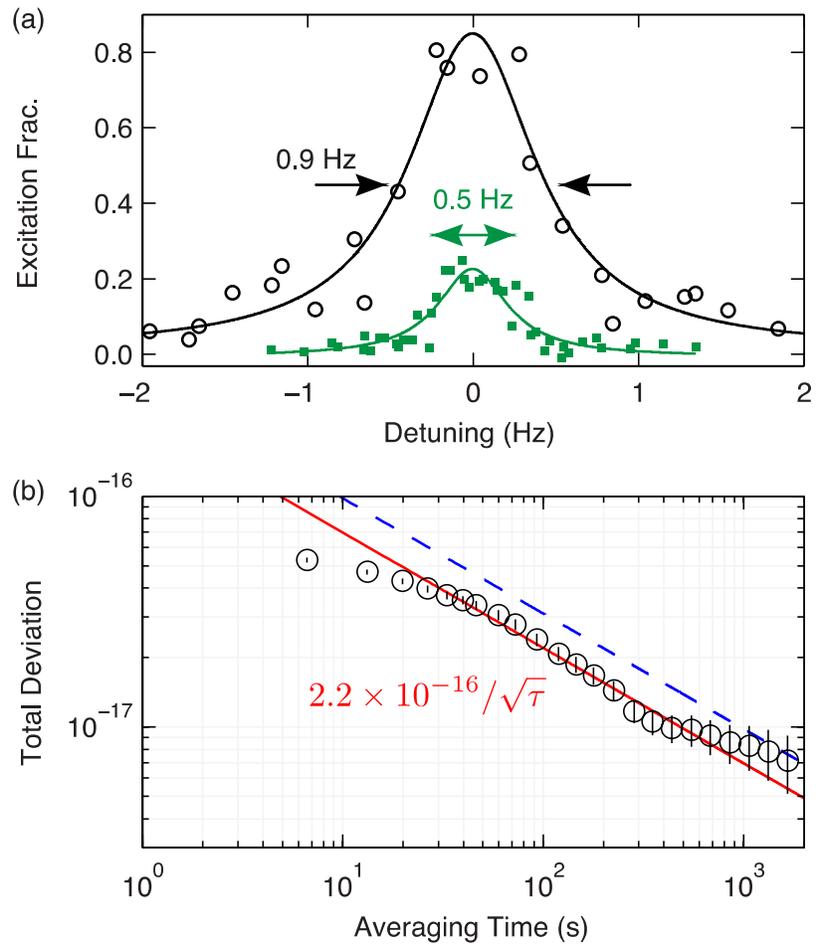

**Figure 1.**

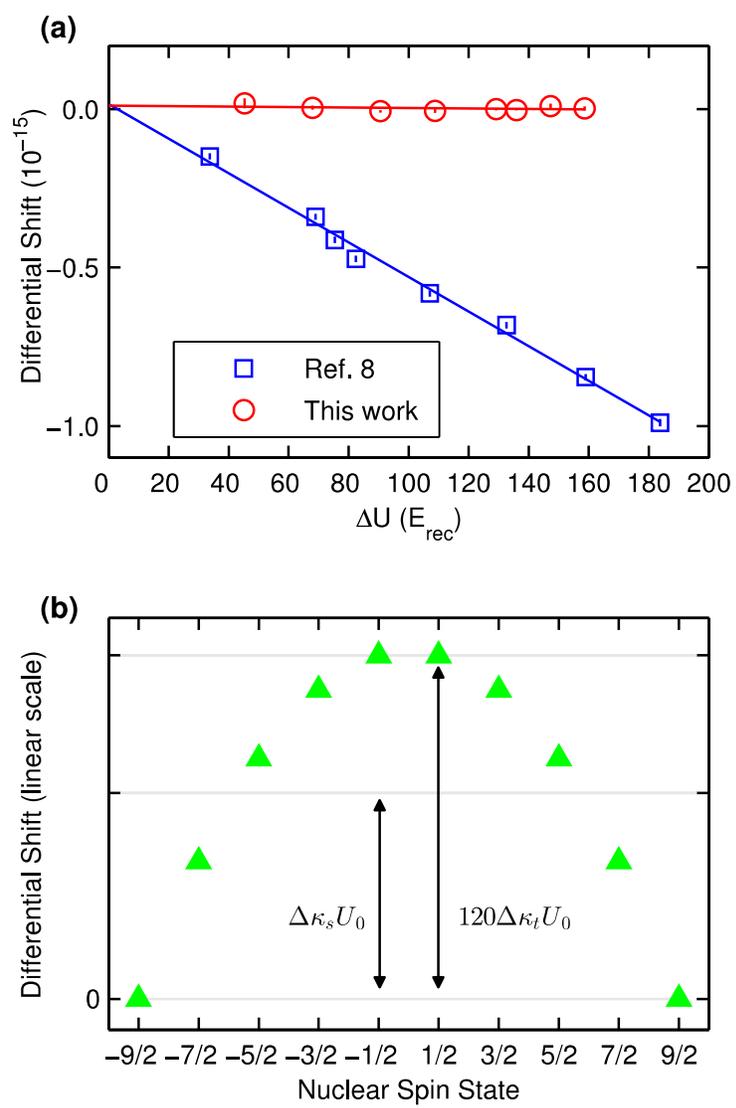

**Figure 2.**

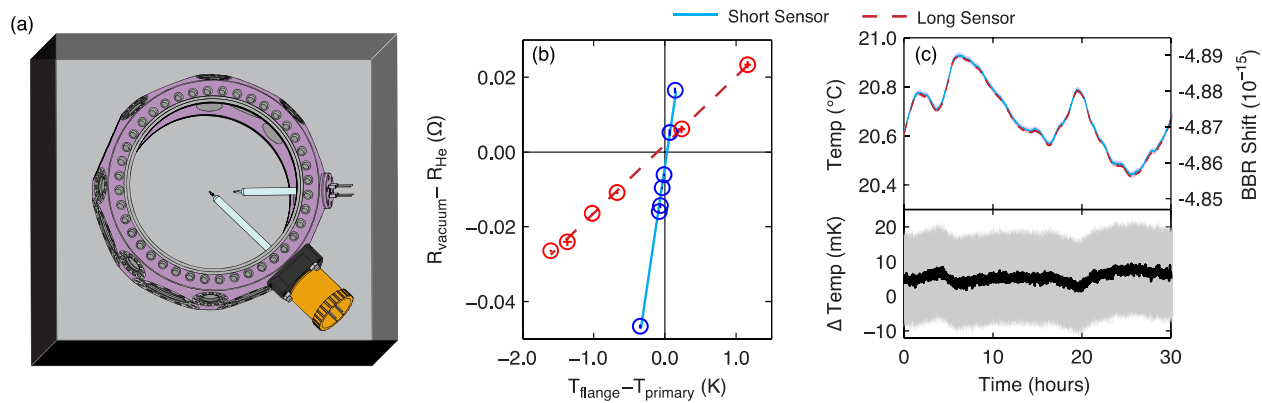

**Figure 3.**

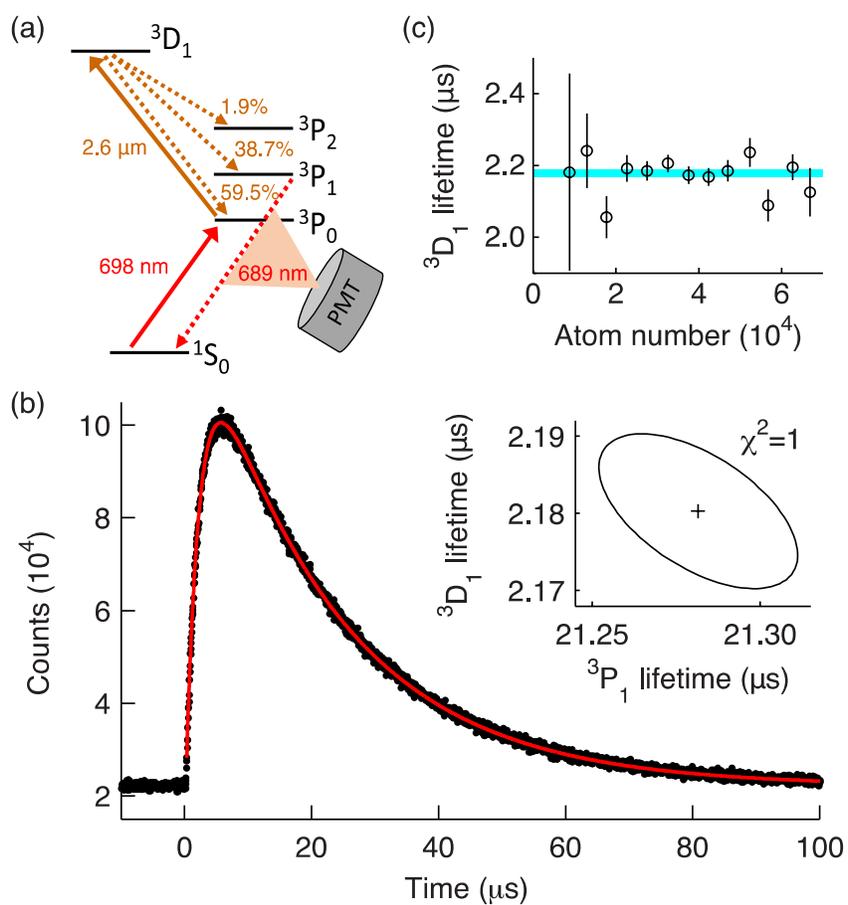

**Figure 4.**

# METHODS

## Sample preparation

We first laser cool a hot strontium beam to 1 mK using a Zeeman slower and 3D magneto-optical trap (MOT) on the $^1S_0 \rightarrow {}^1P_1$ 32 MHz transition at 461 nm. The atoms are further cooled to a few µK with a 3D MOT operating on the $^1S_0 \rightarrow {}^3P_1$ 7.5 kHz intercombination transition at 689 nm. About 2000 atoms are then loaded into a cavity-enhanced 1D optical lattice at 813.4 nm. The cavity mirrors are placed outside the vacuum chamber and the lattice light, generated with an injection-locked Ti:Sapphire laser, is stabilized to the cavity using the Pound-Drever-Hall technique using a double-passed acousto-optic modulation as a frequency actuator.

## Stable laser

The $^{87}$Sr sample is probed on the $^1S_0 \rightarrow {}^3P_0$ 1 mHz clock transition with a 698 nm diode laser, which is stabilized to 26 mHz using a 40 cm Ultralow Expansion glass (ULE) cavity[14,21]. The cavity enclosure features bipolar temperature control, a passive heat shield, a double-chambered vacuum, active vibration cancellation, and acoustic shielding. The stabilized laser passes through an independent acousto-optic modulator (AOM) to steer the frequency of the clock laser light reaching the atoms.

## Atomic servo

The offset of the clock laser frequency relative to the clock transition is determined with Rabi spectroscopy. In this work, measurements utilize Rabi pulse lengths from 160 ms to 4 s. The excited state population fraction after clock spectroscopy is measured by counting the number of $^1S_0$ ground state atoms using $^1S_0 \rightarrow {}^1P_1$ fluorescence, repumping the $^3P_0$ excited state population to the ground state and again counting the number of ground state atoms. To lock the clock laser to the atoms, two excited state population measurements are performed on the clock transition (one on each side of the resonance center). The difference between these measurements is used as an error signal, which is processed by a digital proportional-integral-derivative (PID) controller to steer the laser frequency onto the clock transition resonance.

## Lock in measurements with the atomic servo

Many systematic uncertainties are measured using a digital lock-in technique. In this scheme, an experimental parameter is set at one value, the clock transition is interrogated, and the atomic servo computes a frequency correction[22]. The same procedure is then performed for a different value of the experimental parameter, using a second, independent atomic servo loop. As the experiment alternates between these two states, data is recorded and time stamped. Demodulation occurs in post processing. In all cases we seek the difference between the resonance centers measured by these control loops.

## Density shift

The use of spin-polarized ultracold fermions suppresses s-wave interactions among our atoms; however, p-wave interactions that shift the clock transition can be significant at high precision. This density shift is proportional to the atomic density and insensitive to temperature (due to its p-wave nature and the 1D lattice[18]). The density shift is greatly reduced compared to our previous generation Sr clock due to the use of a cavity-enhanced optical lattice[21]. To measure this shift, we perform a lock-in measurement by modulating the atom number and looking for a frequency shift. Extrapolating this result to an operating atom number of 2000 and trap depth of 71 $E_{rec}$ (where $E_{rec}$ is the lattice photon recoil energy), we reach a density shift of $(-3.5 \pm 0.4) \times 10^{-18}$ (Fig. 5).

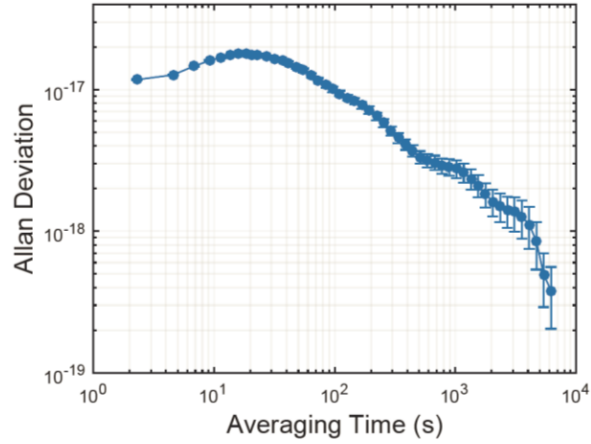

**Figure 5. Evaluation of the density shift.** The overlapping Allan deviation shows the density shift averaging down for 2000 atoms and $U_0$ = 71 $E_{rec}$. The atom number was modulated between 2400 and 12000 atoms.

**Lattice Stark shift**

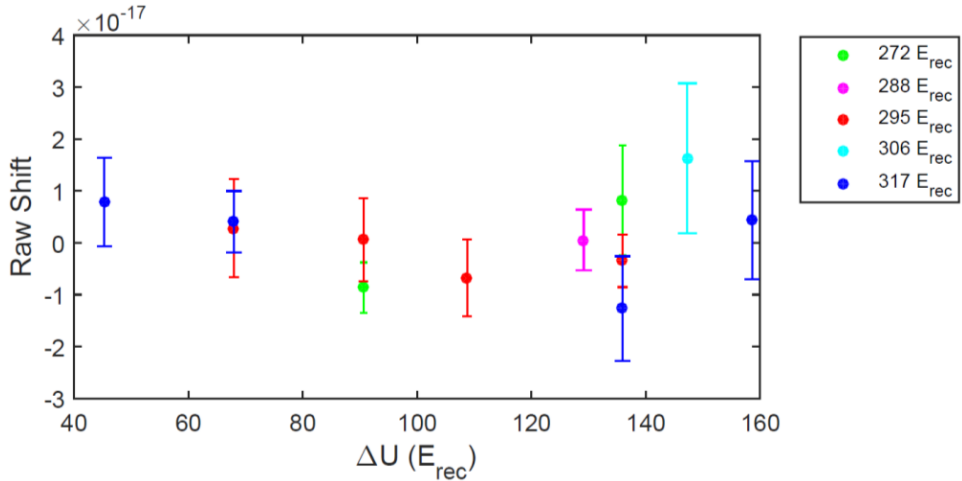

**Figure 6. The lattice Stark data used in this work.** For $\Delta U = U_2 - U_1$, the color coding represents the values of $U_2$ used for each point. Data that has the same value of $\Delta U$ are averaged to produce the points in Fig. 2a.

A lock-in measurement is performed for different lattice powers to study the intensity dependence of the lattice Stark shift. We determine this shift as a function of the optical trap depth at the location of the atoms, $U_0$, which is proportional to the lattice intensity. The value of $U_0$ is determined from the trap frequency along the lattice axis, which is measured using resolved sideband spectroscopy.

Changing $U_0$ also modulates the trap volume, which creates a parasitic density shift that can mimic a lattice light shift. A Gaussian density profile predicts that the density shift scales like $NU_0^{3/2}$. In our system, we experimentally verify this relation with negligible uncertainty. To cancel effects of the density shift on this measurement, we modulate the atom number according to the $NU_0^{3/2}$ scaling such that there is common-mode density shift cancellation. To further ensure that the density shift is removed, in post processing we remove data with the largest atom number fluctuations until the average differential density shift is well below the final measurement precision.

$U_0$ is stabilized with a laser intensity servo by monitoring the cavity lattice transmission. The lattice frequency is locked to a Yb fiber laser optical frequency comb referenced to the NIST maser array. Varying the lattice intensity and frequency, we find the magic wavelength where the clock shift is not responsive to changes in $U_0$.

Drifting background magnetic fields can cause the atom's quantization axis to vary with respect to the clock laser polarization. This creates a drifting ac Stark shift. To solve this problem, we run a background magnetic field servo[1] during the ac Stark shift measurement.

At some level, terms nonlinear in $U_0$ (such as hyperpolarizability and M1-E2 shifts) will be required to precisely model the lattice intensity. To measure these small terms, Ref. 43 relied on the ability to obtain lattices as deep as $10^3$ $E_{rec}$ to achieve a large lattice intensity modulation amplitude. However, the measurement could have been susceptible to technical issues such as noisy tapered amplifier[43] used to generate lattice light or parasitic density shift effects which could be significant for such large changes in lattice trap[25]. To check whether our data supports terms nonlinear in $U_0$ to model the lattice light shift, we use an $F$-test[44] yielding $F = 0.17$ for 22 degrees of freedom (corresponding to unbinned data). Therefore, within our measurement precision, our data only supports a linear model (Fig. 6). We note also that all our lattice Stark shift measurements are made near the clock operating condition, with each data point reaching the statistical uncertainty at the $1 \times 10^{-17}$ level. Together these points determine the Stark shift correction at the $1 \times 10^{-18}$ level for the relevant condition of our clock.

If we were to assume significant hyperpolarizability, we can use our data to infer a hyperpolarizability shift coefficient of $(0.3 \pm 0.3)$ µHz/$E_{rec}^2$. This is consistent with the value reported in Ref. 43. We could also use the hyperpolarizability coefficient of Ref. 43 to correct our data, resulting in a minimal increase in our total uncertainty (from $2.1 \times 10^{-18}$ to $2.4 \times 10^{-18}$). However, since our statistical tests do not justify hyperpolarizability, only linear behavior is assumed in our quoted ac Stark shift.

**Temperature sensors**

The in-vacuum temperature sensors, Heraeus thin-film PRTs, are mounted on the end of borosilicate glass tubes sealed to mini vacuum flanges. PRTs are a well-established technology for accurate thermometry and are ultrahigh-vacuum (UHV) compatible. The PRTs are pre-qualified by cycling their temperatures between an ice melting point (temperature stable to 1 mK) and 200 °C, and then choosing sensors that shifted less than 1 mK over 4 cycles. 4-wire phosphor-bronze connections to the sensors are soldered to electrical feedthroughs in the flanges. The sensor resistance is measured with a bridge circuit, comparing the PRTs to a 1 ppm resistance standard. Resistance measurements are taken with forward and reversed excitation currents for data processing that removes thermocouple effects. Electrical error is quantified in Table II of the main text.

The mounting structures were installed in a test chamber and hand carried on a passenger flight to Gaithersburg, Maryland for calibration at the NIST Sensor Technology Division. At NIST, the sensors were calibrated by comparing them to standard PRTs (SPRTs), traceable to NIST's ITS-90 temperature scale and accurate to 1 mK, using a water comparison bath with 1 mK temperature stability[45]. The temperature uniformity in the isothermal region of the bath is within 1 mK. Since thin-film PRT calibration shifts are quasi-random, mechanisms that could affect the calibrations would cause the two sensors to disagree. Agreement between the sensors throughout the shipping and installation process strongly suggests that no calibration shifts have occurred. Thin-film PRTs are generally robust against calibration shifts due to impacts.

We deal with immersion error by a two-stage process. First, the test chamber is filled with pure helium and the sensors are calibrated to the SPRTs. Data is fit to the Callendar van Dusen equation, $R_{He} = R_0(1 + AT + BT^2)$, where $R_0$, $A$, and $B$ are fit parameters. The helium acts as an exchange gas, enabling radial heat exchange along the glass stem and suppressing immersion error. Second, we measure the sensor resistance under vacuum, $R_{vacuum}$, as a function of $T_{flange} - T_{primary}$. To quantify immersion error, we fit $R_{vacuum} - R_{He} = C(T_{flange} - T_{primary}) + \Delta$, where $C$ and $\Delta$ are fit parameters. These two equations are used to obtain $T_{primary}$ as a function of $R_{vacuum}$ and $T_{base}$. Sensor self-heating is studied by varying the excitation current and extrapolating the results to zero current.

The sensors are installed in the clock chamber using a gas backflow. After installation, sensor baking at 150 °C means that 1.0 mK uncertainty, from thermal cycling, must be added to Table II. One of the sensors can be translated inside the vacuum chamber with an edge-welded bellows. For clock operation, this sensor is positioned 2.5 cm from the atoms to prevent coating with strontium. The temperature difference between the atom location and 2.5 cm away is (1.45 ± 0.03) mK, which is included in Table II.

The sensor translation measurements and temperature measurements throughout the inside of the BBR shield confirm that temperature gradients are small, indicating a well-thermalized environment. Compared to previous efforts[1], temperature gradients in the clock chamber are now smaller because greater care was taken to minimize heat sources inside the BBR shield. To quantify the non-thermal heat shift, we model the geometry and emissivities of the vacuum chamber[1]. We find that our simulation is insensitive to changes in the emissivity values and that the non-thermal heat correction is bounded below the 1 × 10⁻¹⁹ level for our level of temperature uniformity. The non-thermal correction has been included in the "Static BBR" entry of Table I rather than listed in Table II.

**Decay measurement**

After population is driven to the $^3D_1$ state (Fig. 4a), 689 nm fluorescence from the $^3D_1 \rightarrow {}^3P_1 \rightarrow {}^1S_0$ cascade is collected with a photomultiplier tube and then read out and time binned (using a 40 ns bin size) with an SR430 event counter. This photon counting setup provides 0.4 ns of timing uncertainty.

Our statistics have confirmed that the noise in this measurement is Poissonian. Simulating the measurement with the appropriate noise process shows that our fits should be given Poisson weighting to correctly obtain the fit uncertainty.

Other simulations show that the fit does not accrue an appreciable bias due to the specific pulse shape when we use pulses shorter than 300 ns or when we remove data when the pulse is on from the fit. To ensure that this fit bias is doubly suppressed, we take both approaches. We take 0.1 ns as a conservative bound on the remaining uncertainty.

We have calculated the correction due to the $^3D_1$ hyperfine structure to be at the negligible 0.001% level. Therefore, we choose 0.1 ns as a comfortable upper bound on this effect.

We quantify systematic bias from stray distributed-feedback (DFB) laser light by switching off the AOM used to pulse this laser while attempting to scan the $^3P_0 \rightarrow {}^3D_1$ transition. We are able to observe this transition with stray light for exposure times of hundreds of milliseconds. By simulating the results of this scan, we can put a small 0.01 ns upper bound on stray laser light effects. We put the same bound on systematic bias from stray 2.6 μm radiation originating from the ambient heat in our lab.

We study the measured decay rate as a function of atom number to check for density dependence. We confirm that the decay rate is constant in density within our precision using an *F*-test, comparing a constant to a model linear in density. With a value of $F = 0.045$ for the statistic (where there are 11 degrees of freedom), this test indicates no density dependence.

**Dc Stark shift**

A background dc electric field can arise from various sources, such as patch charges[29] or electronics[27]. We have only measured a significant background dc Stark shift along one direction. This axis passes through the two largest viewports and the center of the MOT coils.

To combat possible changes in the dc Stark shift, we actively suppress this shift with electrodes placed on the two large viewports. We measure $v_+$, the total dc Stark shift with the applied field in one direction, and $v_-$, the shift with the applied field flipped in direction. The background field is proportional to ($v_+$- $v_-$), which is processed by a digital Proportional-Integrator servo. The servo applies a voltage to the electrodes to null the background field. The nonlinearity of the shift in electric field means that shift measurements average down rapidly when the background field is well canceled. We measure a low $10^{-20}$ level shift with an uncertainty of (-0.1 ± 1.1) × $10^{-19}$ in 20 minutes of averaging time.

**Probe Stark shift**

We perform this measurement by locking two independent atomic servos to 20 ms and 180 ms π-pulses. By keeping the pulse area, which is proportional to the square root of the probe intensity, fixed at a π-pulse, we can perform low-noise measurements of the probe Stark shift, which is linear in probe intensity. To resolve the shift well, we perform a large amplitude probe intensity modulation using a motor to move a neutral density filter in and out of the clock laser beam path. Control measurements confirm that this filter does not introduce systematic bias.

To prevent issues with many-body effects that might shift the clock transition frequency as a function of atom number, we study the probe Stark shift with a clock operation atom number of 2000. Extrapolating this result to an operating clock pulse of 1 s, we observe a probe Stark shift of (-3.2 ± 1.7) × $10^{-20}$.

**1st-order Zeeman shift**

The 1st-order Zeeman shift is greatly suppressed by averaging locks to the two $m_F = \pm 9/2$ stretched states[26]. A residual 1st-order Zeeman shift could occur if there is appreciable magnetic field drift in between clock interrogations. We combat this by employing active background magnetic field cancellation[1].

The difference between the $m_F = \pm9/2$ stretched state frequency measurements is proportional to the background magnetic field. Drifts in this difference indicate a residual 1$^{st}$-order Zeeman shift. Averaging down this difference, we measure a 1$^{st}$-order Zeeman shift of $(-1.6 \pm 2.0) \times 10^{-19}$.

**2$^{nd}$-order Zeeman shift**

We measure the 2$^{nd}$-order Zeeman shift by monitoring the atomic frequency shift while modulating between high and low bias magnetic field values. We then extrapolate the observed frequency shift to operating conditions, using the fact that the shift is proportional to the bias field squared. The 2$^{nd}$-order Zeeman shift is measured as a function of the frequency difference between the $m_F = \pm 9/2$ stretched states, $\Delta\nu_{stretch}$, which is proportional to the bias field magnitude. For clock operation, $\Delta\nu_{stretch}$ = 300 Hz.

Background field drift can change the direction of the bias field, creating a time varying lattice tensor ac Stark shift that would affect the measurement. To prevent this, we operate a background field cancelation servo. Also, we reduce the sensitivity to drifts by aligning the field and the clock laser polarization. This is done by minimizing $m_F$ changing σ transitions. With this setup, we put a 10$^{-20}$ level upper bound on systematic bias from field drift.

We measure the 2$^{nd}$-order Zeeman shift coefficient, the shift normalized by $\Delta\nu^2_{stretch}$, to be $(-5.82 \pm 0.07) \times 10^{-16}/kHz^2$. This number is an atomic property and is independent of a particular measurement, so we average this result with 4 other determinations of this coefficient[1,25,46,47]. The final value for the shift at $\Delta\nu_{stretch}$ = 300 Hz is $(-51.7 \pm 0.3) \times 10^{-18}$. We use a reduced chi square $\chi^2_{red}$ inflated uncertainty to account for non-statistical variations between these data points.

**Other shifts**

Line pulling occurs when off-resonant spectroscopic features can slightly shift the clock transition frequency. This can be caused by imperfect spin polarization leaving population in $m_F$ states aside from ±9/2, clock laser ellipticity causing us to drive $m_F$-changing σ transitions, or clock transition sidebands that result from tunneling between lattice sites. Calculations and data allow us to put a conservative upper bound on this effect at $1 \times 10^{-19}$.

The 1$^{st}$-order Doppler effect is not present in an optical lattice probed along the lattice axis, where the optical phase of the lattice and that of the clock probe lasers are referenced to a common mirror. A 2$^{nd}$-order Doppler shift is, in principle, present, but it is estimated to be at the 10$^{-21}$ level. We put a comfortable $1 \times 10^{-19}$ bound on this effect.

Collisions with the background gases in our UHV vacuum chamber can shift the clock transition frequency. At normal operating vacuum pressure, the background gas is largely hydrogen. We use the model of Ref. 48 to put an upper bound on this effect of $6 \times 10^{-19}$.

Steady-state error in the atomic servo could shift the measured clock transition frequency. We average lock data and find a servo offset of $(-5 \pm 4) \times 10^{-19}$.

Clock operation utilizes an AOM to scan the frequency and pulse the intensity of the clock laser. Phase transients occurring when this AOM pulses would appear as frequency shifts in clock measurements. We study the AOM phase transients by looking at the beat of the -1$^{st}$ AOM order with the 0$^{th}$-order on a digital phase detector. We also calibrated the phase transients of the detector itself. Drawing on the analysis of Ref. 49, we infer an AOM phase chirp shift of $(6 \pm 4) \times 10^{-19}$.

**Statistical methods**

To calculate the shift of a given record, we perform a post processing demodulation of the data to extract a signal. The shift represents the mean of this signal. The statistical uncertainty is calculated from the standard deviation of the mean. If the reduced chi square $\chi^2_{red} > 1$, the statistical uncertainty is inflated by $\sqrt{\chi^2_{red}}$. To remove the effects of residual laser drift, which is highly linear, from lock-in measurements, we use "3-point strings." This analysis involves processing successive triplets of frequency measurements in linear combinations meant to cancel linear drift[50].